\documentclass[preprint,12pt]{article}
 
\usepackage{amssymb}
\usepackage{caption}
\usepackage{breqn}
\usepackage{graphicx}

\usepackage{caption}
\usepackage{subcaption}
\usepackage{physics}
\usepackage[numbers, sort&compress]{natbib} 

\begin{document}

\textbf{Pomeron Evolution and Squeezed States in Quantum Optics}
\begin{center}

\small
 Dima Cheskis    and Alex Prygarin   
\\
  Department of Physics, Ariel University, Ariel, Israel 
\end{center}
\normalsize

\normalsize

\begin{abstract}
We apply the formalism of  coherent states in quantum optics  to pomeron evolution and show that 
   evolving squeezed  pomeron states are equivalent to   pomeron  fan  diagrams at the leading order of   perturbative expansion. Based on our results, we interpret the action of the displacement operator as pomeron propagation and the action of the squeeze operator as 
   pomeron interaction. 

\end{abstract}

\section{Introduction}
 
The squeezed coherent states are used in quantum optics and other fields for reducing uncertainty of physical observables like coordinate or momentum. The high energy scattering by pomerons 
is described in QCD by Balitsky-Fadin-Kuraev-Lipatov~(BFKL)~\cite{BFKL1, BFKL2, BFKL3, BFKL4, BFKL5} approach and has no obvious connection to the field of quantum optics expect general concepts of wave mechanics, scattering amplitude, opacity etc.   Inspired 
by recent developments in using basic concepts of quantum  information science for the high energy scattering~\cite{Kharzeev:2017qzs, Kharzeev:2021nzh,Hentschinski:2022rsa, Kutak:2023cwg,Chachamis:2023omp,Hentschinski:2024gaa, Datta:2024hpn, Afik:2025ejh},
we apply the formalism of the squeezed coherent states to the evolution of pomerons as a proper degree of freedom in the  high energy scattering. The  squeezed coherent states in quantum optics 
are introduced using the displacement operator and the squeeze operator~\cite{Walls:1983zz}. 
The  displacement operator is   defined by 
\begin{eqnarray}\label{disop}
\hat{D}(\beta) =e^{\beta \hat{a}^\dagger-\bar{\beta} \hat{a} },
\end{eqnarray}
where $\beta $ is a complex parameter  and $\hat{a} | n\rangle=C_+(n) | n+1\rangle$ and $ \hat{a}^\dagger | n\rangle=C_{-}(n) | n-1\rangle$ are the ladder operators satisfying the commutation relation 
\begin{eqnarray}
[\hat{a}, \hat{a}^\dagger]=1.
\end{eqnarray}
 The displacement operator is the unitary operator that shifts the ladder operator by a complex constant
$\hat{D}^\dagger (\beta) \hat{a} \hat{D}(\beta) = \hat{a}+\beta $. Applied to the vacuum, it produces the coherent state $ \hat{D}(\beta) | 0 \rangle =  | \beta \rangle$.
 It can be written in another form  using Baker-Campbell-Hausdorff relation as follows
\begin{eqnarray}\label{disop}
\hat{D}(\beta) =e^{\beta  
\hat{a}^\dagger-\bar{\beta} \hat{a} }=e^{-\frac{1}{2} |\beta|^2} 
e^{\beta \hat{a}^\dagger }
e^{-\bar{\beta}  \hat{a}}.
\end{eqnarray}
The squeeze operator is  given by  
\begin{eqnarray}\label{squeezeop}
\hat{S}(z)=e^{\frac{1}{2} z \hat{a}^\dagger \hat{a}^\dagger- \frac{1}{2}\bar{z} \hat{a}  \hat{a} },
\end{eqnarray} 
where $z$ is a complex parameter. The squeezed state is produced applying $\hat{S}(z)$ an $\hat{D}(\beta) $ to vacuum
\begin{eqnarray}\label{squeezedstate}
 \hat{S}(z) \hat{D}(\beta)  |  0 \rangle = | \beta, z\rangle.
\end{eqnarray}

In the case of the quantum harmonic oscillator with energy $E_n=\hbar \omega \left( n+\frac{1}{2} \right)$
the ladder operators are defined by 
\begin{eqnarray}\label{aadaggerxp}
\hat{a}=\sqrt{\frac{m \omega}{2 \hbar}} \hat{x} +i\frac{\hat{p}}{ \sqrt{ 2 \hbar m \omega}}, \;\; \hat{a}^\dagger=\sqrt{\frac{m \omega}{2 \hbar}} \hat{x} -i\frac{\hat{p}}{ \sqrt{ 2 \hbar m \omega}}
\end{eqnarray}
and for real $z$ the uncertainty in  $x$ and $p$ for the squeezed state are given by 
\begin{eqnarray}\label{deltaxp}
\left( \Delta x  \right)^2 = \frac{\hbar}{2 m \omega  } e^{-2 z}, \;\;
\left( \Delta p  \right)^2 = \frac{\hbar m \omega }{2 } e^{+2 z} 
\label{deltaxdeltap}
\end{eqnarray}
resulting in 
\begin{eqnarray}\label{uncertrel}
\left( \Delta x  \right) \left( \Delta p  \right)=\frac{\hbar}{2}
\end{eqnarray}
This illustrates the fact that the uncertainty relation $\left( \Delta x  \right) \left( \Delta p  \right) \geq \frac{\hbar}{2}$ is saturated by the squeezed states. The main advantage of the squeezed states is  that  the
individual uncertainties of $\Delta x$ and $\Delta p$ can be controlled by the parameter $z$ keeping their product constant. 

It is instructive to rewrite the uncertainty relation in terms of the ladder operators $\hat{a}$ and $\hat{a}^\dagger$  noting 
that\footnote{Let $\mathcal{H}$ be a Hilbert space over $\mathbb{C}$ and $\hat{A} \in  \mathbb{B}(\mathcal{H})$ be a bounded linear operator. 
Then the imaginary part of  $\hat{A}$ is the Hermitian operator $ \Im ( \hat{A})\equiv     
\frac{1}{2 i }\left(
\hat{A}-\hat{A}^\dagger \right)$. In a similar way we define the real part as $ \Re( \hat{A}) \equiv   
\frac{1}{2  }\left(
\hat{A}+\hat{A}^\dagger \right)$. }
\begin{eqnarray}
\hat{x}=\frac{2 \hbar }{\gamma} \Re (\hat{a})=\frac{2 \hbar }{\gamma}\frac{1}{2}\left(
\hat{a}+\hat{a}^\dagger \right), \;\; \hat{y}= \gamma  \Im ( \hat{a})= \gamma  
\frac{1}{2 i }\left(
\hat{a}-\hat{a}^\dagger \right),
\label{imreala}
\end{eqnarray}
where $\gamma =\sqrt{ 2 m \omega \hbar}$.

  Then in accordance with \eqref{deltaxdeltap} we write the eigenvalues of $\Re (a) $ and $\Im (a) $ as 
\begin{eqnarray}
\left( \Delta \Re (a)  \right)^2=  4 e^{-2 z }, \;\;
\left( \Delta \Im (a)  \right)^2= \frac{1}{4} e^{+2 z }
\end{eqnarray}
and finally  \begin{eqnarray}
\left( \Delta \Re (a)  \right) \left( \Delta \Im (a)  \right)=1.
\end{eqnarray}
 
In the next section, we apply the squeezed coherent states formalism to the evolution of pomerons in the high energy scattering. We show that the squeezed states  at the leading order  of the perturbative expansion represent the 
fan diagrams with pomeron splitting through triple pomeron vertex.

\section{Pomeron evolution}

In this paper we discuss the case of zero transverse momentum that allows a simple and transparent analogy between pomeron evolution picture and the development of the  squeezed coherent states. Nevertheless, in order to make the model meaningful we must include some information from the   full-dimensional model. For example, in the inspiring paper on the dipole model~\cite{Mueller} the author  included the relative minus sign between two terms $\hat{a}^\dagger\hat{a}^\dagger \hat{a}$ and $\hat{a}^\dagger\hat{a} $ based the
two dimensional loop integration (see eq.~(15), eq.~(37) and eq.~(38) in \cite{Mueller}). 
In order to have a  realistic model, 
the  commutation relation  $ [\hat{a}, \hat{a}^\dagger]=1$ should be supplemented with  momentum conservation, which avoids non-physical combinations of the ladder operators, where all momenta are incoming $\hat{a} \hat{a} \hat{a}$ or  outgoing $\hat{a}^\dagger \hat{a}^\dagger \hat{a}^\dagger$.
This condition arises from the transverse momentum conservation in the $t$-channel and, in the perturbative picture, is related to the polarization of the 
high energy $s$-channel gluons.
  
  We also define "smallness" of the ladder operators based on our knowledge from the perturbative QCD. 
Let us define some small parameter $\epsilon$, which is related  to $\alpha_{QCD}$ in the perturbative regime and write  the smallness of ladder operators  by their expectation values 
\footnote{Strictly speaking, the smallness of the linear bounded operator can be defined through its operator norm.}
\begin{eqnarray}\label{Oaadagger}
\langle
\hat{a} \rangle \propto \mathcal{O}(\epsilon),\;\;\; \langle
\hat{a}^\dagger \rangle \propto \mathcal{O}(1/\epsilon) 
\end{eqnarray}
so that the expectation value of the  product $ \hat{a}^\dagger  \hat{a}$ of two bounded operators of order of unity, i.e. $\mathcal{O}(\epsilon^0)$.
Note that the definition of smallness in \eqref{Oaadagger} leaves the  smallness of the expectation value of their commutation relation $[\hat{a}, \hat{a}^\dagger]$ unchanged. 
In other words, we use a freedom of rescaling the fields and the accompanying constants in the individual terms of the Lagrangian in the Regge Field Theory leaving the overall smallness of those terms unchanged. 
For the sake of clarity of presentation, we do emphasize this point further in the text, referring to operators as "small" with definite smallness in terms of powers of $\epsilon$.

We use  the initial condition where the vacuum pomeron state  $| 1\rangle$ has one pomeron. 
Expanding the product of the displacement and the squeeze operator we obtain
 \begin{eqnarray}
 \hat{S}(z) \hat{D}(\beta)=\sum_{k=0}^\infty \sum_{m=0}^\infty \frac{\left( \frac{1}{2} z \hat{a}^\dagger \hat{a}^\dagger- \frac{1}{2} \bar{z} \hat{a} \hat{a}  \right)^k}{k!} \frac{ \left( \beta \hat{a}^\dagger- \bar{\beta} \hat{a}  \right)^m}{m!} 
 \end{eqnarray}
Assigning the relative smallness to the parameters in the displacement and the squeeze operator 
\begin{eqnarray}\label{Obetaz}
|\beta| \propto \mathcal{O}(\epsilon^2), \;\;\; |z| \propto \mathcal{O}(\epsilon^3)
\end{eqnarray}

we extract the leading term of  the squeeze operator $\hat{S}(z)$ in powers of   $\epsilon$
\begin{eqnarray}
\hat{S}(z) \to \left( \frac{1}{2} z \hat{a}^\dagger \hat{a}^\dagger- \frac{1}{2} \bar{z} \hat{a} \hat{a}  \right)^k \simeq  \frac{1}{2^k} z^k \left( \hat{a}^\dagger \hat{a}^\dagger \right)^k
\end{eqnarray}
At the first sight the same approximation could be done with $\left( \beta \hat{a}^\dagger- \bar{\beta} \hat{a}  \right)^m$, but we must impose 
momentum conservation condition where all annihilation operators $\hat{a}$ are meant to have incoming momenta whereas the creation operators $\hat{a}^\dagger$ are defined for outgoing momenta. In other words, any combination of the ladder operators of the same type only, whether it is $\hat{a} \hat{a} \hat{a}$ or $\hat{a}^\dagger \hat{a}^\dagger \hat{a}^\dagger$ vanishes by virtue of the momentum conservation. Despite the fact that we are dealing with  a toy model of zero transverse dimension, the energy-momentum conservation cannot be ignored and must be implicitly used for eliminating the non-physical solutions.  
In addition, any term of the type 
\begin{eqnarray}
\frac{\frac{1}{2^k} z^k \left( \hat{a}^\dagger \hat{a}^\dagger \right)^k}{k!} \frac{\left(- \beta^{m-1}\bar{\beta} \right) \left( \hat{a}^\dagger\right)^{m-1} \hat{a}  }{m!} 
\end{eqnarray}
is suppressed compared to the follwing term
\begin{eqnarray}\label{leading}
\frac{\frac{1}{2^k} z^k \left( \hat{a}^\dagger \hat{a}^\dagger \right)^k}{k!} \frac{\left(\bar{\beta}\right)^m  \hat{a}^{m}  }{m!}.
\end{eqnarray}
For example, we take term
$\hat{a}^\dagger \hat{a}^\dagger \hat{a}^\dagger \hat{a}^\dagger  \hat{a} \hat{a}
$, 
which would correspond to annihilation of two pomerons and production of four pomerons, i.e. two subsequent pomeron splittings. 
The powers of  $\hat{a}^\dagger$ may come either from the expansion of $\hat{D}(\beta)$ or $\hat{S}(z)$. Let us consider the case where all powers of $\hat{a}^\dagger$ are due expansion of $\hat{S}(z)$ and all powers of $\hat{a}$ are from the expansion of $\hat{D}(\beta)$
\begin{eqnarray}\label{big}
-\frac{\frac{1}{2^2} z^2  \hat{a}^\dagger \hat{a}^\dagger \hat{a}^\dagger \hat{a}^\dagger  }{2!} \frac{\left(\bar{\beta}\right)^2  \hat{a} \hat{a}  }{2!} \simeq \mathcal{O}(\epsilon^8),
\end{eqnarray}
which directly follows from \eqref{Oaadagger} and \eqref{Obetaz}. The same $ \hat{a}^\dagger \hat{a}^\dagger \hat{a}^\dagger \hat{a}^\dagger \hat{a} \hat{a}$ structure can be obtained by  having $\hat{a}^\dagger \hat{a}^\dagger $ from $\hat{S}(z)$ and 
$   \hat{a}^\dagger \hat{a}^\dagger \hat{a} \hat{a}$ expanding $\hat{D}(\beta)$, namely
\begin{eqnarray}\label{small}
 \frac{1}{2} z  \hat{a}^\dagger \hat{a}^\dagger     \frac{-4 \left(\bar{\beta}\right)^2 \beta^2   \hat{a}^\dagger \hat{a}^\dagger  \hat{a} \hat{a}  }{4!} \simeq \mathcal{O}(\epsilon^9).
\end{eqnarray}
One can see that the  expression in \eqref{small} is suppressed with respect to a similar expression in \eqref{big}. 
Analogous situation happens for any expansion where at least one power of $\hat{a}^\dagger \hat{a}^\dagger$ comes from $\hat{D}(\beta)$, those terms have  meaning of sub-leading corrections in perturbative QCD.  

In accordance with smallness considerations and the momentum conservation condition, for each power of a couple of the  ladder operators  $\hat{a}^\dagger \hat{a}^\dagger$ there must be at least one power of $\hat{a}$ 
 \begin{eqnarray}\label{newleading}
\sum_{k=1}^\infty \sum_{m=1}^\infty \frac{\frac{1}{2^k} z^k \left( \hat{a}^\dagger \hat{a}^\dagger \right)^k}{k!} \frac{\left(\bar{\beta}\right)^m  \hat{a}^{m}  }{m!} \delta_{k,m} 
=
\sum_{k=1}^\infty   \frac{\frac{1}{2^k} \left( z \bar{\beta} \right)^k \left( \hat{a}^\dagger \hat{a}^\dagger   \right)^k \hat{a}^k}{(k!)^2}   
\end{eqnarray}

Note that $\left(\hat{a}^\dagger \hat{a}^\dagger \right)^k  \hat{a}^{k}$ does not equal $\left(\hat{a}^\dagger \hat{a}^\dagger \hat{a} \right)^k $ due to the fact that $\hat{a}$ and $\hat{a}^\dagger$ do not commute. 
For example, consider
\begin{eqnarray}
&& \left. \frac{\frac{1}{2^k} \left( z \bar{\beta} \right)^k \left( \hat{a}^\dagger \hat{a}^\dagger   \right)^k \hat{a}^k}{(k!)^2}\right|_{k=2}
= 
\frac{\frac{1}{2^2} \left( z \bar{\beta} \right)^2 \hat{a}^\dagger \hat{a}^\dagger \hat{a}^\dagger \hat{a}^\dagger \hat{a} \;\hat{a} }{(2!)^2}  \nonumber
\\
&=&
\frac{\frac{1}{2^2} \left( z \bar{\beta} \right)^2
\left( \hat{a}^\dagger \hat{a}^\dagger \hat{a}   \right)
\left( \hat{a}^\dagger \hat{a}^\dagger \hat{a}   \right)   }{(2!)^2}-2\frac{\frac{1}{2^2} \left( z \bar{\beta} \right)^2
 \left( \hat{a}^\dagger \hat{a}^\dagger  \hat{a}^\dagger \hat{a}   \right)   }{(2!)^2}   
\end{eqnarray}
where the last term is negligible compared to the previous terms in the expansion
\begin{eqnarray}
 \frac{\frac{1}{2} z \beta \bar{\beta}  
 \left( \hat{a}^\dagger \hat{a}^\dagger  \hat{a}^\dagger \hat{a}   \right)   }{2!} 
 \end{eqnarray}
 due to  different   smallness    of $z$ and $\beta$,i.e. $\left( z \bar{\beta} \right)^2 \propto \mathcal{O}(\epsilon^{10})$ versus $ z \beta \bar{\beta} \propto \mathcal{O}(\epsilon^{7})$. The subsequent application $k$-times of the three operator product  $ \hat{a}^\dagger \hat{a}^\dagger \hat{a} $ is related to the $k$-th power of this operator by 
 \begin{eqnarray}
(\hat{a}^\dagger \hat{a}^\dagger \hat{a})(\hat{a}^\dagger \hat{a}^\dagger \hat{a})... (\hat{a}^\dagger \hat{a}^\dagger \hat{a})= k! (\hat{a}^\dagger \hat{a}^\dagger \hat{a})^k 
 \end{eqnarray}
 due to the fact that $\hat{a}$ and $\hat{a}^\dagger$ do not commute. 
 Plugging this back into \eqref{newleading} we obtain
 \begin{eqnarray}\label{newleading}
&&\sum_{k=0}^\infty \sum_{m=0}^\infty \frac{\frac{1}{2^k} z^k \left( \hat{a}^\dagger \hat{a}^\dagger \right)^k}{k!} \frac{\left(\bar{\beta}\right)^m  \hat{a}^{m}  }{m!} \delta_{k,m} 
=
\sum_{k=0}^\infty   \frac{\frac{1}{2^k} \left( z \bar{\beta} \right)^k \left( \hat{a}^\dagger \hat{a}^\dagger   \right)^k \hat{a}^k}{(k!)^2}   \nonumber \\
&&  
\simeq \sum_{k=0}^\infty   \frac{\frac{1}{2^k} \left( z \bar{\beta} \right)^k \left( \hat{a}^\dagger \hat{a}^\dagger  \hat{a} \right)^k }{k!}    =e^{\frac{1}{2} z \bar{\beta} \hat{a}^\dagger \hat{a}^\dagger  \hat{a}  }.
\end{eqnarray}

In a similar way we consider the  product of two operators  $\hat{a}^\dagger \hat{a} $. Any power of  that type can be only obtained from the expansion of the displacement operator $\hat{D}(\beta)$ in \eqref{disop}. Any power of $\hat{a}$ or $\hat{a}^\dagger$ that comes from the squeeze operator $\hat{S}(z)$ in \eqref{squeezeop} will be suppressed by virtue of the relative smallness of $z$ and $\beta$ in \eqref{Obetaz}.
Thus from \eqref{disop} we have the leading contribution to the bilinear form $\hat{a}^\dagger \hat{a} $ as follows
\begin{eqnarray}\label{leadaadagger}
&& \sum_{k=0}^\infty  \sum_{m=0}^\infty  \frac{\left(\beta \hat{a}^\dagger     \right)^k}{k!} \frac{\left(-\bar{\beta} \hat{a}      \right)^m}{m!} \delta_{m,k}=\sum_{k=0}^\infty     \frac{\left(-\beta  \bar{\beta}\right)^k \left( \hat{a}^\dagger      \right)^k \hat{a}^k}{(k!)^2}
\nonumber  
\\
&& \simeq  \sum_{k=0}^\infty     \frac{\left(-\beta  \bar{\beta}\right)^k \left( \hat{a}^\dagger      \hat{a} \right)^k  }{k! } =e^{- |\beta|^2 \hat{a}^\dagger      \hat{a} }
\end{eqnarray}

 Using the expressions in \eqref{newleading} and \eqref{leadaadagger} we write the leading contribution from the product of the displacement operator $\hat{D}(\beta)$ and  the squeeze operator $\hat{S}(z)$ to the squeezed state defined in \eqref{squeezedstate}
 \begin{eqnarray}\label{squeezedstate2}
 \hat{S}(z) \hat{D}(\beta)  |  0 \rangle  \simeq  e^{\frac{1}{2} z \bar{\beta} \hat{a}^\dagger \hat{a}^\dagger  \hat{a}  }  e^{- |\beta|^2 \hat{a}^\dagger      \hat{a} }  |  0 \rangle 
\end{eqnarray}
We use  the Baker–Campbell–Hausdorff formula $e^X e^Y= e^Z$ 
\begin{eqnarray}
Z=X+Y +\frac{1}{2} \left[  X, Y \right] +
\frac{1}{12} \left[  X, \left[  X, Y \right] 
 \right]+\frac{1}{12} \left[  Y, \left[  Y, Y \right] 
 \right]+...
\end{eqnarray}
to rewrite the lhs in \eqref{squeezedstate2} as an exponent of the sum of the two operators. 
The commutator of those two operators reads
\begin{eqnarray}\label{commXY}
  [ z \bar{\beta}\hat{a}^\dagger \hat{a}^\dagger \hat{a}, |\beta|^2  \hat{a}^\dagger \hat{a}  ]= z|\beta|^2 \bar{\beta}\hat{a}^\dagger \hat{a}^\dagger \hat{a}
\end{eqnarray}
It   is strongly suppressed
compared to 
\begin{eqnarray}
\frac{1}{2} z \bar{\beta} \hat{a}^\dagger \hat{a}^\dagger  \hat{a}
\end{eqnarray}
due to the fact that 
$
 z \bar{\beta} |\beta|^2  \propto \mathcal{O} (\epsilon^9)
$
compared to 
$ z \bar{\beta}    \propto \mathcal{O} (\epsilon^5).
$
Other commutator terms in the Baker–Campbell–Hausdorff formula are also suppressed and can be neglected in the leading order. 
This results into 
 \begin{eqnarray}\label{squeezedstate3}
 \hat{S}(z) \hat{D}(\beta)  |  0 \rangle  \simeq 
  e^{\frac{1}{2} z \bar{\beta} \hat{a}^\dagger \hat{a}^\dagger  \hat{a}  - |\beta|^2 \hat{a}^\dagger      \hat{a} } |  0 \rangle .
\end{eqnarray}
Note the both terms in the exponent are of the same order in the small parameter $\epsilon$
\begin{eqnarray}
\frac{1}{2} z \bar{\beta} \hat{a}^\dagger \hat{a}^\dagger  \hat{a} \propto \mathcal{O}(\epsilon^4), \;\;  
 \;\; |\beta|^2 \hat{a}^\dagger      \hat{a} \propto \mathcal{O}(\epsilon^4).
\end{eqnarray}

Finally, we compare this with the preparation of a state with fan pomeron diagrams  
in the zero transverse dimension~\cite{Mueller}. The multi-pomeron state is given by 
\begin{eqnarray}\label{expV1V2}
e^{Y (V_1+V_2)} \hat{a}^\dagger |  0 \rangle,  
\end{eqnarray}
where $Y$ is the rapidity. The vertices $V_1$ and $V_2$ are given by the triple pomeron vertex 
\begin{eqnarray}\label{V1}
V_1= \alpha \;\hat{a}^\dagger \hat{a}^\dagger \; \hat{a}
\end{eqnarray}
and the term responsible for the pomeron propagation 
\begin{eqnarray}\label{V2}
V_2= -\alpha  \; \hat{a}^\dagger \; \hat{a}
\end{eqnarray}
Using the multi-pomeron state in \eqref{expV1V2} one can define the generating function 
\begin{eqnarray}\label{ZYu}
Z(Y,u)= \langle 0 | e^{\hat{a} u}      e^{Y (V_1+V_2)} \hat{a}^\dagger |  0 \rangle,  
\end{eqnarray}
with initial condition 
\begin{eqnarray}\label{init}
Z(Y=0, u)=u
\end{eqnarray}
and the property 
\begin{eqnarray}
Z(Y, u=1)=1,
\end{eqnarray}
which reflects the conservation of the total probability. 

Differentiating  $Z(Y, u)$ with respect to rapidity $Y$ leads to differential equation 
\begin{eqnarray}
\frac{1}{\alpha} \frac{d Z}{d Y} = -Z+Z^2.
\end{eqnarray}
Its is solved  by 
\begin{eqnarray}
Z(Y,u)= \frac{u}{u +(1-u)e^{\alpha Y}}= \sum_{n=1}^\infty P_n(Y) u^n,
\end{eqnarray}
where the probabilities $P_n(Y)$ read
\begin{eqnarray}\label{Pn}
P_n(Y)= e^{-\alpha Y} \left( 1- e^{-\alpha Y}  \right)^{n-1}. 
\end{eqnarray}
Using the probability $P_n(Y)$ one can calculate the mean multiplicity of produced particles  and entropy relating the high energy evolution model to the experimental data and basic concepts quantum information science~\cite{Kharzeev:2017qzs, Kharzeev:2021nzh,Hentschinski:2022rsa, Kutak:2023cwg,Chachamis:2023omp,Hentschinski:2024gaa, Datta:2024hpn, Afik:2025ejh}.

If we impose that the initial condition for the squeezed state in  \eqref{squeezedstate3} consists of one particle we can write it as 
\begin{eqnarray}\label{squeezedstate4}
 \hat{S}(z) \hat{D}(\beta)  \hat{a}^\dagger |  0 \rangle  \simeq 
  e^{\frac{1}{2} z \bar{\beta} \hat{a}^\dagger \hat{a}^\dagger  \hat{a}  - |\beta|^2 \hat{a}^\dagger      \hat{a} }  \hat{a}^\dagger  |  0 \rangle.
\end{eqnarray}

Comparing the expressions in \eqref{squeezedstate4} and \eqref{expV1V2} we identify the displacement and the squeezing parameters as follows 
\begin{eqnarray}\label{ztoY}
\frac{1}{2} z \bar{\beta} \epsilon = \alpha Y, \;\; |\beta|^2= \alpha Y,
\end{eqnarray}
where we rescale back the ladder operators in such a way,  that they are of the same order in the small parameter $\epsilon$ 
\begin{eqnarray}
\hat{a}^\dagger \propto \mathcal{O}(\epsilon^0), \;\;\; \hat{a} \propto \mathcal{O}(\epsilon^0).
\end{eqnarray} 
Plugging \eqref{ztoY} into \eqref{squeezedstate3} we obtain the leading order  \textit{squeezed pomeron state}  
\begin{eqnarray}\label{squeezedpomeron}
 \hat{S}(z) \hat{D}(\beta) \hat{a}^\dagger  |  0 \rangle  \simeq 
  e^{\alpha Y \hat{a}^\dagger \hat{a}^\dagger  \hat{a}  - \alpha Y \hat{a}^\dagger      \hat{a} } \hat{a}^\dagger  |  0 \rangle 
\end{eqnarray}
that corresponds to the pomeron fan diagrams described by the generating function in \eqref{ZYu}. 

It is worth emphasizing that the term $\alpha \hat{a}^\dagger      \hat{a}$ in the exponent of \eqref{squeezedpomeron} represents the propagation of pomerons and it originates merely from the expansion of the displacement operator $\hat{D}(\beta)$, while the term $\alpha \hat{a}^\dagger \hat{a}^\dagger  \hat{a}$ comes from cross term  expansion of  the displacement operator $\hat{D}(\beta)$ and the squeezing operator $\hat{S}(z)$.  This fact allows for a rather transparent interpretation of the displacement operator $\hat{D}(\beta)$ as the operator responsible for the propagation of the pomerons and the squeezing operator $\hat{S}(z)$ as the operator responsible for the triple pomeron interaction. 
The squeezed state $\hat{S}(z) \hat{D}(\beta) \hat{a}^\dagger |  0 \rangle $  in \eqref{squeezedpomeron} is the state that includes all possible pomeron propagation and splitting suggesting  that  the meaning of \textit{pomeron squeezing} is a related to the   unitarization of the scattering amplitude. The suppressed terms that  we neglected in the course of the derivation of \eqref{squeezedpomeron} 
would account for pomeron loops.  This presents the main result of our study.

\section{Pomeron splitting versus photon squeezing}
Relating the pomeron interaction to the squeezed states of photons one should keep in mind that there is a major difference between the two. Namely, in contrast to the pomerons, the photons do not interact with each other belonging to $U(1)$ group interaction. Second order processes can result in one photon absorption accompanied by two photon emission, which effectively can be viewed as one photon being split into two photons. As we showed, mathematically this is equivalent to the pomeron splitting through triple pomeron vertex. 

The pomeron splitting implies the momentum of the parent pomeron being divided between two daughter pomerons. In terms of the squeezed photon states, those states  are experimentally prepared using a resonant cavity with frequency $\omega_0$ and semitransparent mirrors that partially reflect photons at frequencies $\omega_0$ and $2\omega_0$. The incoming photon at frequency $2\omega_0$ is annihilated with simultaneous creation of two  photons at  frequency $\omega_0$, mimicking photon splitting.  The cavity may have two resonant frequencies $\omega_1$ and $\omega_2$ that lead to creation of two photons at those two frequencies once an incoming  photon having frequency $\omega_0=\omega_1+\omega_2$ is annihilated. 

\section{Conclusion and Discussions}\label{}
 In this paper we applied the formalism of the  squeezed  coherent states  to the pomeron evolution in zero transverse momentum. We  showed that the pomeron evolution can be described as an evolving squeezed state of pomeron  in the leading order of the perturbative expansion. We associate the action of the displacement operator with the pomeron propagation and the squeeze operator as the operator responsible for the pomeron interaction.

\section{Acknowledgement}\label{}

We are indebted to Sergey Bondarenko  for inspiring discussions on the topic. 
This work is supported in part by "Program of HEP support- Council of Higher Education of Israel".



\begin{thebibliography}{00}







\bibitem {BFKL1} 
L.~N.~Lipatov,
"Reggeization of the Vector Meson and the Vacuum Singularity in Nonabelian Gauge Theories"
   Sov. J. Nucl. Phys. 
{\bf 23} (1976) 338; 
\bibitem {BFKL2} 
V.~S.~Fadin, E.~A.~Kuraev and L.~N.~Lipatov,
"On the Pomeranchuk Singularity in Asymptotically Free Theories",  Phys. Lett.
\textbf{B 60} (1975) 50;
\bibitem {BFKL3} 
E.~A.~Kuraev, L.~N.~Lipatov and V.~S.~Fadin,
"Multi - Reggeon Processes in the Yang-Mills Theory",
  Sov. Phys. JETP \textbf{44} (1976) 443;
\textbf{45} (1977) 199;
\bibitem {BFKL4} 
Ya.~Ya.~Balitskii and
L.~N.~Lipatov, 
"The Pomeranchuk Singularity in Quantum Chromodynamics", 
Sov. J. Nucl. Phys. \textbf{28} (1978) 822



\bibitem{BFKL5} 
  V.~S.~Fadin and L.~N.~Lipatov,
  ``BFKL pomeron in the next-to-leading approximation,''
  Phys.\ Lett.\ B {\bf 429}, 127 (1998)
  doi:10.1016/S0370-2693(98)00473-0
  [hep-ph/9802290].
  

\bibitem{Kharzeev:2017qzs}
D.~E.~Kharzeev and E.~M.~Levin,
``Deep inelastic scattering as a probe of entanglement,''
Phys. Rev. D \textbf{95}, no.11, 114008 (2017)
doi:10.1103/PhysRevD.95.114008
[arXiv:1702.03489 [hep-ph]].
  
  
\bibitem{Kharzeev:2021nzh}
D.~E.~Kharzeev,
``Quantum information approach to high energy interactions,''
Phil. Trans. A. Math. Phys. Eng. Sci. \textbf{380}, no.2216, 20210063 (2021)
doi:10.1098/rsta.2021.0063
[arXiv:2108.08792 [hep-ph]].




\bibitem{Hentschinski:2022rsa}
M.~Hentschinski, K.~Kutak and R.~Straka,
``Maximally entangled proton and charged hadron multiplicity in Deep Inelastic Scattering,''
Eur. Phys. J. C \textbf{82}, no.12, 1147 (2022)
doi:10.1140/epjc/s10052-022-11122-1
[arXiv:2207.09430 [hep-ph]].


\bibitem{Kutak:2023cwg}
K.~Kutak,
``Entanglement entropy of proton and its relation to thermodynamics entropy,''
[arXiv:2310.18510 [hep-ph]].



\bibitem{Chachamis:2023omp}
G.~Chachamis, M.~Hentschinski and A.~Sabio Vera,
``Von Neumann entropy and Lindblad decoherence in the high-energy limit of strong interactions,''
Phys. Rev. D \textbf{109}, no.5, 054015 (2024)
doi:10.1103/PhysRevD.109.054015
[arXiv:2312.16743 [hep-th]].

\bibitem{Hentschinski:2024gaa}
M.~Hentschinski, D.~E.~Kharzeev, K.~Kutak and Z.~Tu,
``QCD evolution of entanglement entropy,''
Rept. Prog. Phys. \textbf{87}, no.12, 120501 (2024)
doi:10.1088/1361-6633/ad910b
[arXiv:2408.01259 [hep-ph]].



\bibitem{Datta:2024hpn}
J.~Datta, A.~Deshpande, D.~E.~Kharzeev, C.~J.~Na\"\i{}m and Z.~Tu,
``Entanglement as a Probe of Hadronization,''
Phys. Rev. Lett. \textbf{134}, no.11, 111902 (2025)
doi:10.1103/PhysRevLett.134.111902
[arXiv:2410.22331 [hep-ph]].

\cite{Afik:2025ejh}
\bibitem{Afik:2025ejh}
Y.~Afik, F.~Fabbri, M.~Low, L.~Marzola, J.~A.~Aguilar-Saavedra, M.~M.~Altakach, N.~A.~Asbah, Y.~Bai, H.~Banks and A.~J.~Barr, \textit{et al.}
``Quantum Information meets High-Energy Physics: Input to the update of the European Strategy for Particle Physics,''
[arXiv:2504.00086 [hep-ph]].




\bibitem{Walls:1983zz}
D.~F.~Walls,
``Squeezed states of light,''
Nature \textbf{306}, 141-146 (1983)
doi:10.1038/306141a0





\bibitem{Mueller}
A.~H.~Mueller,
``Unitarity and the BFKL pomeron,''
Nucl. Phys. B \textbf{437}, 107-126 (1995)
doi:10.1016/0550-3213(94)00480-3
[arXiv:hep-ph/9408245 [hep-ph]].
  
  
  
  
  
  
  
  
  
  
  

  


  
\end{thebibliography}
\end{document}